\documentclass[review,authoryear,number]{elsarticle}

\usepackage{graphicx}
\usepackage[latin2]{inputenc}
\usepackage{multirow}
\usepackage{lineno}

\journal{Journal of Atmospheric and Solar-Terrestrial Physics}

\begin{document}


\begin{frontmatter}

\title{Hot flow anomaly remnant in the far geotail?}

\author[fmi,ggi]{G.~Facsk{\'o}}
\corref{cor}
\cortext[cor]{Corresponding author}
\ead{facsko.gabor@csfk.mta.hu}
\author[irap,estec]{A.~Opitz}
\author[irap]{B. Lavraud}
\author[ssl]{J.~G.~Luhmann}
\author[berkeley]{C.~T.~Russell}
\author[irap]{J-A. Sauvaud}
\author[irap]{A. Fedorov}
\author[ggi]{A.~Kis}
\author[ggi]{V.~Wesztergom}
\address[fmi]{Finnish Meteorological Institute, Earth Observation Unit, Helsinki, Finland}
\address[ggi]{Geodetic and Geophysical Institute, Research Centre for Astronomy and Earth Sciences, HAS, Sopron, Hungary}
\address[irap]{IRAP (CNRS-UPS), University of Toulouse, Toulouse, France}
\address[estec]{ESTEC, ESA, Noordwijk, The Netherlands}
\address[ssl]{Space Sciences Laboratory, University of California, Berkeley, USA}
\address[berkeley]{Institute of Geophysics and Planetary Physics, University of California, Los Angeles, USA}

\begin{abstract}
A hot flow anomaly (HFA) like event was observed by the Solar TErrestrial RElations Observatory (STEREO) in the night side magnetosheath in the far tail in February-March 2007. The magnetic signature of the tangential discontinuity was visible, but the resolution of the plasma ion data is not sufficient for our analysis, so a method is given to identify HFAs without solar wind velocity measurements. The event observed in the night side magnetosheath in the far tail might be the remnant of an HFA event, a not-so-active current sheet. This observation suggests that the lifetime of the HFAs might be several 10 minutes, much longer than the expected several minutes.
\end{abstract}

\begin{keyword}
hot flow anomaly; tangential discontinuity; magnetotail; STEREO
\end{keyword}

\end{frontmatter}


\section{Introduction}
\label{sec:intro} 

Hot flow anomalies (HFAs, also known as hot diamagnetic cavities and as active current sheet) were discovered in the 1980s \citep{schwartz85, thomsen86:_hot}. A tangential discontinuity (TD) interacts with the bow shock and in the meantime the convective electric field on at least one side points toward the current sheet \citep{schwartz00:_condit_earth,facsko08:_clust, facsko09:_clust,facsko10:_study_clust}. The particles are accelerated and turned back from the quasi-perpendicular bow-shock and the electric field focuses them to the current sheet \citep{thomsen86:_hot}. The TD leads the particles back to the bow shock so they can gain large energies. After the intersection region of the TD and the bow shock reaches the quasi-parallel region a beam is ejected and its particles interact with the unperturbed solar wind flow \citep{kecskemety06:_distr_rapid_clust,omidi07:_format}. This interaction forms a diamagnetic cavity by Alfv\'en waves and heats the plasma inside the cavity \citep{thomas89:_three}. The two particle populations are visible in the young and in the proto-HFAs. Actually, they unify soon and the event is then called mature HFA \citep{lucek04:_clust, tjulin08:_wave, zhang10:_time_histor_event_macros_inter_subst}. All terrestrial HFAs \citep[and the Venusian HFAs in][]{collinson12:_hot_flow_anomal_venus} have been observed near the subsolar point or on the dayside as opposed to the events observed in Martian and Kronian systems \citep{oeieroset01:_hot_martian,masters08:_cassin_satur,masters09:_hot_satur}. A very extent HFA was (or two simultaneous events were) observed by \citet{safrankova12:_asymm}.

Here we analyse STEREO Behind (STB) measurements when the STB was situated beyond the Earth orbit, entered the tail after the STB second Moon flyby. The structure of this paper is as follows: we describe the observation in Section~\ref{sec:obs}, analyse the event in Section~\ref{sec:tail}, discuss the results in Section~\ref{sec:disc}; finally the conclusions are given in the last Section~\ref{sec:summ}. 

\section{Observations}
\label{sec:obs}

A series of criteria was set for the selection of HFA events  \citep{facsko08:_clust, facsko09:_clust, facsko10:_clust_hot_flow_anomal_obser, facsko10:_study_clust, zhang10:_time_histor_event_macros_inter_subst,wang12:_cases_hot_flow_anomal_clust,kovacs13:_turbul}:
\begin{enumerate}
\item The rim of the cavity must be visible as a sudden increase of magnetic field magnitude compared to the unperturbed solar wind value. Inside the cavity the magnetic field magnitude drops and its direction turns around. \label{enum:b}
\item The solar wind speed drops, and its direction always turns away from the Sun-Earth line. \label{enum:drop}
\item The solar wind temperature increases and its value reaches up to several ten million degrees. \label{enum:t}
\item The solar wind particle density also increases at the rim of the cavity and drops inside the HFA. \label{enum:n}
\item The presence of a TD in the upstream magnetic field. \label{enum:td1}
\item The angle of the TD normal and the solar direction must be greater than $45^o$.  \label{enum:td2}
\item The convective electric field vectors (-vxB) must po\-int toward the TD on at least one side of the discontinuity (preferably on both sides). \label{enum:vxb}
\item Energetic ions appear often but not always. When suprathermal particles are detected with 28-410\,keV energy then the increased level of the flux starts before the magnetic signatures and ends after them. 
\item Long wavelength wave activity can be observed inside the cavity and the plasma is turbulent inside.  \label{enum:turbu}
\item The solar  wind speed is higher than the average (about 600\,km/s). \label{enum:high}
\end{enumerate} 
We had to neglect some of these criteria, namely (\ref{enum:drop}) and (\ref{enum:high}), because the plasma ion instrument was not switched on before the Moon flyby and its resolution was not enough to provide solar wind speed data during the tail event. 

HFAs were observed and simulated without the presence of a TD \citep{omidi13:_spont,zhang13:_spont}, so it seems that the conditions (5) and (6) are obsolete. Furthermore a large amount of HFAs without focusing convective electric field was observed \citep{wang13:_clust_feb,wang13:_clust_jan,wang13:_hot}, so the role of the convective electric field in HFA formation became uncertain, so criteria (7) does not seem necessary. Beside these new features we had to restrict our survey for the classic HFA events, because of technical problems. In the studied intervals the STEREO SWEA and PLASTIC instruments had no data or its accuracy was insufficient. All data were generated manually. The intervals of manual data production were selected by the magnetic field configuration (5,6,7). Involving the Spontaneous HFAs and events where (7) criteria is not satisfied would have increased significantly the amount of data production. So we restricted our survey for the classic HFA events.

We use 8\,Hz resolution STEREO mag\-ne\-to\-me\-ter \citep[MAG,][]{acuna08:_stereo_impac_magnet_field_exper}, one-minute resolution ion plasma \citep[PLASTIC,][]{galvin08:_plasm_suprat_ion_compos_plast} and 2\,s or 30\,s resolution electron temperature and density data \citep[SWEA,][]{sauvaud08:_impac_solar_wind_elect_analy_swea,fedorov11:_impac_solar_wind_elect_analy_swea,opitz10:_tempor_evolut_solar_wind_elect}. The instruments of STEREO spacecraft were designed for working in the solar wind and studying the heliosphere. The 8\,Hz temporal resolution (32\,Hz in burst mode) of MAG is sufficient for detecting the diamagnetic cavity but, since the typical length of an HFA event is 2--3 minutes, the PLASTIC plasma instrument maximal temporal resolution of one-minute provides only one-two points inside the cavity. This fact made us base the event identification on magnetic measurements and  electron temperature and density observations. The plasma ion velocity measurements are only needed for checking two of the ten conditions listed above. 

The STEREO spacecraft were launched on a large eccentricity orbit and after a Moon flyby, the two space observatories were separated. Before the critical manoeuvre with our moon the two satellites crossed the Earth magnetosphere numerous times. After their separation, STB intersected the flank, the magnetosheath, the magnetopause and the regions of the night side magnetosphere in the early phase of its heliospheric mission \citep{kistler10:_escap_o,sauvaud11:_far_r_e,opitz14:_solar_stereo}. The Earth flybys  and the near-Earth period of STEREO were analysed. Several suspicious events were found during the STEREO near Earth phase in the January--April, 2007 interval, however finally only one event in the night side magnetosheath in the far tail was analysed.

\section{HFA candidates in the far tail}
\label{sec:tail}

\begin{figure}[h]
\vspace*{2mm}
\begin{center}
\includegraphics[width=200pt]{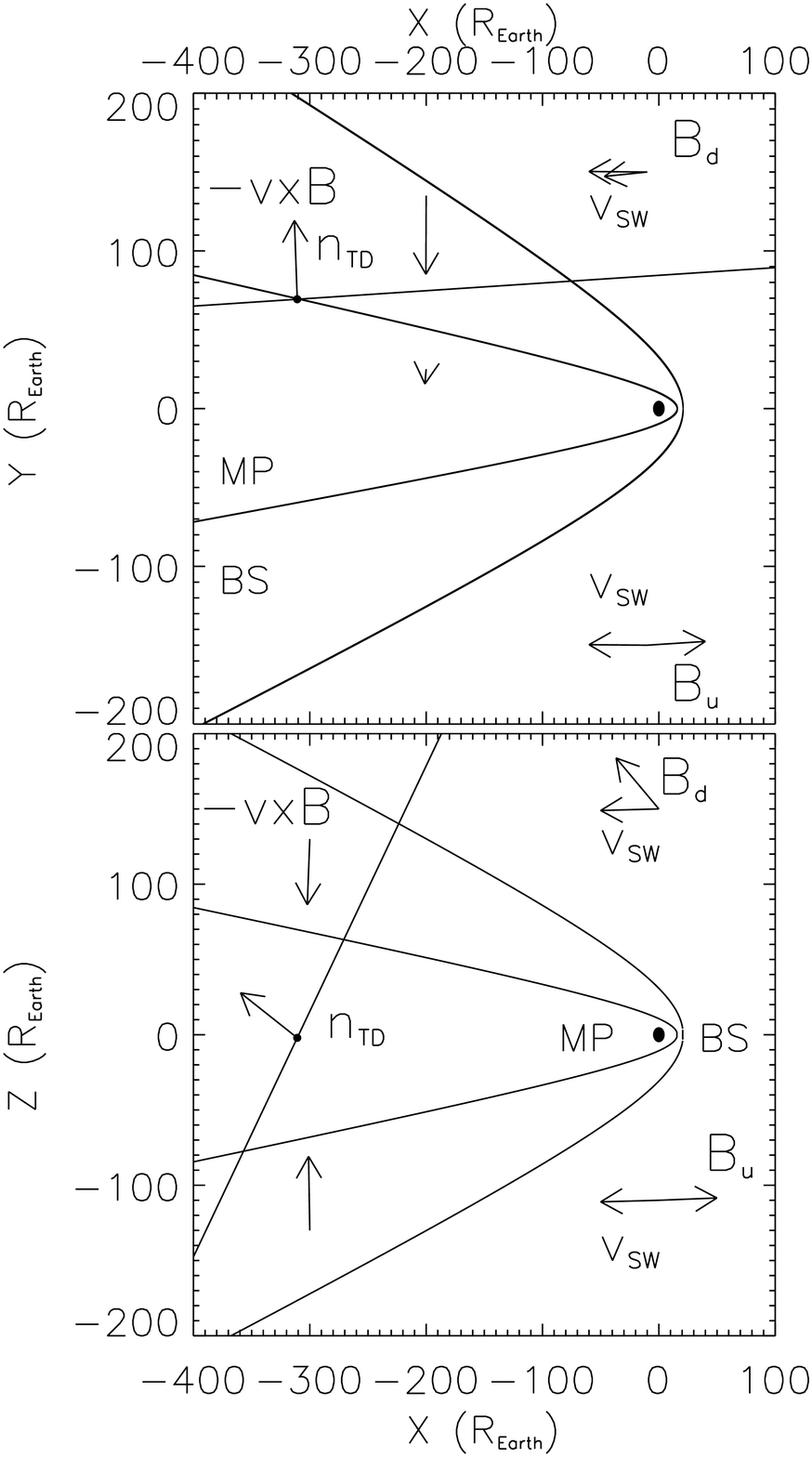}
\end{center}
\caption{The configuration of the HFA event from 00:35 to 00:50 on March 1, 2007 in GSE system where $n_{TD}$ is the discontinuity normal, $B_{u}$, $B_{d}$ are the magnetic fields situated upstream and downstream from the discontinuity, $v_{SW}$ is the solar wind velocity and $-v\times B$ is the convective electric field. The TD normals are situated at the spacecraft position. The bow shock based on the \citet{peredo95:_three_alfven_mach} model and an arbitrary magnetopause were drawn on the figure.}
\label{fig:tailmag}
\end{figure}

\begin{figure*}[t]
\vspace*{2mm}
\begin{center}
\includegraphics[width=350pt]{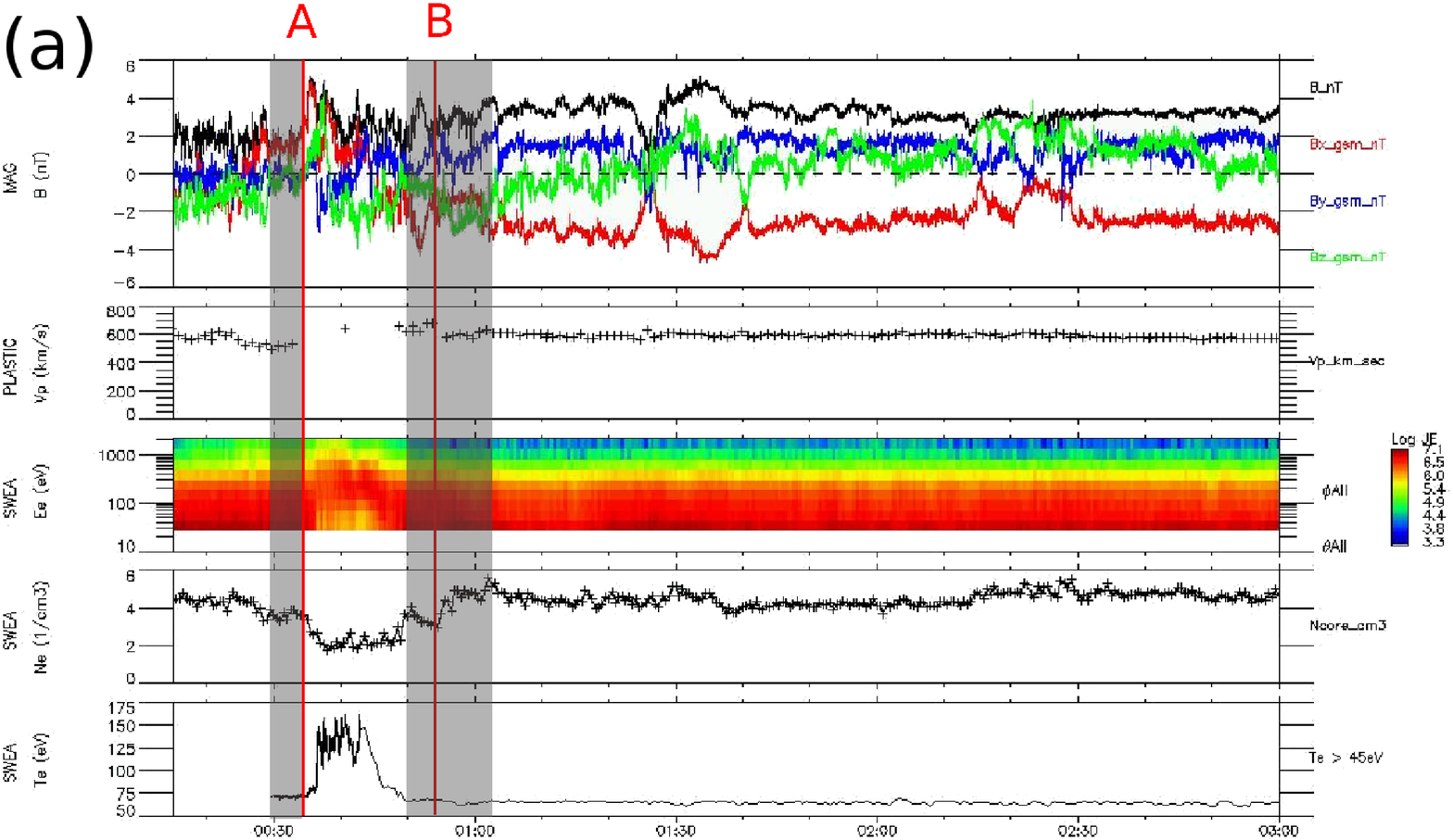}
\includegraphics[width=350pt]{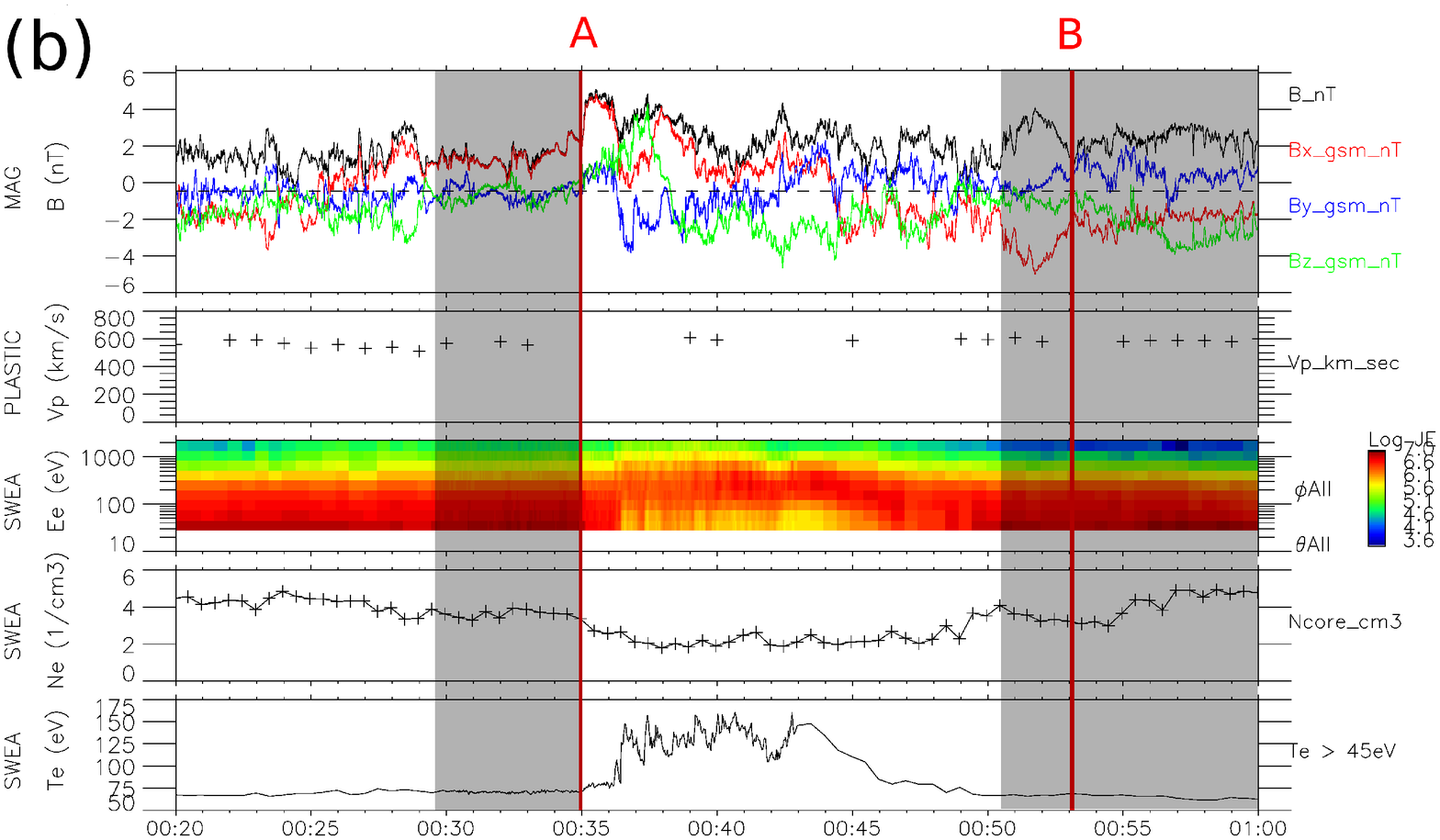}
\end{center}
\caption{STEREO-B in-situ measurements of the HFA candidate event (between A and B red vertical lines) observed from 00:35 to 00:50 on 1st March 2007. Plot (a) gives an overview about the event, the magnetosheath and magnetopause crossing, Plot (b) is a zoom into the event. The same measurements were drawn on both plots. Upper panel: IMPACT MAG magnetic field. Second panel: PLASTIC proton bulk velocity. Third to fifth panels: IMPACT SWEA electron energy spectrogram, core electron density \citep{opitz10:_tempor_evolut_solar_wind_elect,fedorov11:_impac_solar_wind_elect_analy_swea} and suprathermal electron temperature, respectively. Minimum variance and cross product methods were applied for the gray intervals.}
\label{fig:tailplot}
\end{figure*}

From January 1 to April 30, 2007, the orbit of STB spacecraft was situated close to the geotail (Figure~\ref{fig:tailmag}). HFA events candidates were searched in the time interval using the 8\,Hz temporal resolution IMPACT/MAG magnetometer measurements and 2\,s or 30\,s resolution SWEA electron temperature and density measurements. From March 1 to April 30, the one-min resolution PLASTIC plasma measurements were also analysed. A good candidate was found between 00:35-00:52 (UT) March 1, 2007. On March 1 at 00:35 (UT) the STEREO B spacecraft was situated in the following position in the GSE system: 
\begin{eqnarray*}
\label{eq:norm2}
r_{STB} &=& (-310.8, 69.3, -2.1)\,R_{E}. 
\end{eqnarray*}
The TD normal was determined with the minimum variance and the cross product methods in GSE system. The cross product method was applied using the intervals of the minimum variance method. The minimum variance method proves whether the discontinuity is tangential. Then the TD normal is calculated using the more accurate cross product method. These methods were applied for the 00:29:54.5-00:35:09.5 and 00:49:56.5-01:02:42.5 (UT) intervals on March 1, 2007. The result of the cross product vector calculation is: 
\begin{eqnarray*}
\label{eq:tdnorm2}
n_{TD} &=& (-0.05, 0.99,  0.03).
\end{eqnarray*}
(see the grey fields in Figure~\ref{fig:tailplot}). This is a tangential discontinuity because the magnetic field is almost zero in the minimum variance system, the ratio of the eigenvalues is $\lambda_2 / \lambda_3 = 2.1$, furthermore the direction of the minimum variance eigenvector and the cross product vector is almost the same (Condition~\ref{enum:td1}). The calculated electric field points to the discontinuity on both sides of the TD in 3D so Condition~\ref{enum:vxb} is fulfilled.

The drop in the magnetic field and in the density indicates the presence of a cavity (Conditions~\ref{enum:b} and \ref{enum:n}, Figure~\ref{fig:tailplot}). The STEREO SWEA and PLASTIC measurements confirm the existence of the cavity too. The density drop is visible in SWEA measurements as well as the slight temperature increase (Condition~\ref{enum:t}). The data gap and the ion spectrum indicates a peculiar event \citep[as is in][for Venus an HFA observation]{collinson12:_hot_flow_anomal_venus}, however this fact emphasizes that the velocity distribution is non-Maxwellian there, other\-wise fitting for the distribution functions and so the calculation of the moments would be possible and easier. The density drop and the temperature increase are visible in the SWEA electron measurements. The IMPACT instrument suite has an automatic control of measurement mode. If it observes a notable event, it switches to high resolution (2\,s) mode. The SWEA burst mode with 2-second resolution indicates here a more special event than magnetosheath or tail crossing. During this event the solar wind speed was around 600\,km/s, higher than the average (Condition~\ref{enum:high}). This condition was not part of the selection criteria. The survey was based on magnetic field and electron data. However in this case the moments could be calculated using PLASTIC measurements.

\section{Discussion}
\label{sec:disc}
 
This event has several unusual features. First of all, it is longer (15\,minutes) than a typical HFA, that lasts only several minutes \citep{facsko09:_clust,facsko10:_study_clust}. The magnitude of the magnetic field and the electron density is less than in the environment. This drop is not so significant in the magnetic field. The rims of the cavity does not show a strong shock at all. No sign of the rim is visible in the electron density, however it is usually visible only in the ion density \citep[see for example in][Figure~6]{facsko10:_study_clust}. The temperature increases significantly, however it is not several million degree as expected based on Condition~\ref{enum:t}. This event has HFA like features, but it cannot be stated that this event is an HFA based on these observations.

The direction of the TDs can also be determined in the solar wind using the combination of minimum variance and cross product methods. In that case, the core, the middle of the event must not be cut, because it is not so noisy and disturbed like in the case of a hot diamagnetic cavity. For those discontinuities, the expected value of the eigenvalue ratio is 8, or greater. In this case, in these highly disturbed events this ratio could be only 2-3, even if the cavity is not considered. Despite of this quite low value, the event must be considered as TD, because the higher value is rare in such turbulent region \citep[][Table~2]{facsko09:_clust}. So we can argue that this is a TD, or a TD related event.

The event was observed close to the magnetopause, that is considered a TD by definition: the solar wind does not enter to the obstacle, but it flows around the terrestrial magnetosphere \citep{baumjohann96:_basic}. So the normal component of the magnetic field and solar wind velocity vanish, there is no mass flow through the magnetopause. However the layer between the magnetopause and the magnetosphere is not always TD as expected. Furthermore after crossing the magnetopause the density drops, the magnetic field magnitude does not change significantly and electron temperature also drops. The ion temperature can be similar or higher, but all changes occur in the same time. 

The foreshock events have different classes: foreshock cavities \citep{fairfield90:_upstr}, foreshock cavitons \citep{blanco-cano09:_global}, Larmor radius size density holes \citep{parks06:_larmor_earth} and Spontaneous HFAs \citep[SHFA,][]{zhang13:_spont,omidi13:_spont}. Actually the presence of the TD and the increased temperature excludes all of these events; the temperature reaches several million degrees in the Spontaneous HFAs, but TD cannot be observed nearby, that is why this feature was named to \textit{Spontaneous} HFA.

A possible explanation of the observed magnetic signature would be the presence of a plasmoid. When a plasmoid passes, the $B_z$ component of the magnetic field changes its sign and helicity can be seen in the magnetic field related to a flux rope. Furthermore the density increases in the far tail plasmoid observations \citep[see][Figure~2,~5, Figure~3 and Figure~11, respectively]{kiehas12:_earth_artem,li13:_plasm_artem,voeroes14:_winds_condit_ram_co_ram}. The density in the plasmoids does not change \citep{ieda98:_statis_geotail} or increases \citep{moldwin92:_isee_geotail} when the plasmoids move downtail. The recent plasmoid observations in the far tail ($255\,R_{E}$) by STEREO also saw increased density \citep[][Figure~3]{sauvaud11:_far_r_e}. Here the magnetic field configuration also changes and helical structure can be observed, because a TD-like structure was passed. Against plasmoids the electron density drops here, so the measured lower electron density excludes plasmoids as explanation.

Ions could escape from the magnetopause and increase both of the electron and ion temperatures significantly. The ratio of the ion and electron temperatures ($T_i/T_e$) could reach high value (10-15) close to the magnetopause \citep{wang12:_spatial}. This significant temperature incrementation appears on the dusk side of the magnetopause based on simulations and observations \citep{gkioulidou11:_effec,wang12:_spatial}. The STEREO PLASTIC and SWEA instruments provide ion and electron velocity distribution functions from significantly higher energy than the THEMIS \citep{angelopoulos08:_themis_mission} electrostatic analyser \citep[ESA;][]{mcfadden08:_themis_esa_plasm_instr_in_calib} and solid state telescope (SST). So these temperatures are not equivalent to the temperatures used by \citet{wang12:_spatial}. Furthermore the PLASTIC cannot provide temperature during the events, so the  $T_i/T_e$ cannot be calculated. However the ion and electron spectra show possible temperature incrementation and this event is situated on the dusk side of the magnetosphere. But a TD is present during the event and the density drops significantly, so the events could occur because of leaking magnetopshere ions, but the probability of an HFA remnant presence is higher.

Several possible explanations were excluded a\-bo\-ve. We argue that, here we detected the remnant of an HFA event. So this is not an HFA, but it was an HFA once. Here we saw only a current sheet, that was once active. The acceleration processes were stopped, there is no sign of the expansion any more. The magnetic configuration reminds us for an HFA cavity, the TD is still visible. The density is lower in the remnant of the cavity and the plasma still hotter than the magnetosheath and the lobe. This is a super-mature, or a dead HFA.

\section{Summary and conclusions}
\label{sec:summ}

The Earth orbit phase and the beginning of the STE\-RE\-O mission shortly after the lunar flyby were analysed in this study. A new method was introduced to try identifying HFAs without plasma ion measurements. This way of identification might be useful for analysing for example Venus Express or MESSENGER data. Many conditions of HFA formation were fulfilled as we saw in Section~\ref{sec:tail}, but the event is not an HFA -- any more. We argue that we observed an event that might have been HFA once. In this case the first terrestrial observation in the night side magnetosheath in the far tail was presented. The event was observed very far from the subsolar point of the bow shock ($318\,R_{E}$ downtail). The HFA-like phenomenon observed in the tail is really unexpected because it suggests that the HFAs can exist and expand a long time $\left(\sim50^m\right)$ after leaving the formation region. Naturally the typical signature become weaker and weaker as the cavity expands and vanishes. Alternative explanation could be that the particle acceleration continues (or it might restart) in the tail. Further hybrid simulations and observation studies are necessary to explain this experience properly.

\paragraph{Acknowledgement} Figure~\ref{fig:tailplot} was created by the CL tool so we thank E. Penou at IRAP for software support. This work was supported by the OTKA Grant K75640 of the Hungarian Scientific Research Fund and the ECLAT EU FP7 Project, the Grant Agreement number is 263325. Andrea Opitz is currently a Research Fellow at ESA ESTEC. {\'A}rp{\'a}d Kis was supported by the OTKA Grant PD78674 of the Hungarian Scientific Research Fund. G{\'a}\-bor Facsk{\'o} acknowledges Heli Hietala for useful discussions and Anna-M\'aria V{\'\i}gh for improving the English of the paper.

\bibliographystyle{elsarticle-harv}
\bibliography{jastp-2013-stereo-tx}  

\end{document}